# Formation of a White-Light Jet within a Quadrupolar Magnetic Configuration

Boris Filippov[1] • Serge Koutchmy[2] • Ehsan Tavabi[3]

[1]*Pushkov Institute of Terrestrial Magnetism, Ionosphere and Radio Wave Propagation, Russian Academy of Sciences, Troitsk, Moscow Region 142190, Russia*
*(e-mail:* bfilip@izmiran.ru*)*

[2]*Institut d'Astrophysique de Paris, CNRS and Univ. P.& M. Curie, 98 bis Boulevard Arago, 75014 Paris, France*
*(e-mail:* koutchmy@iap.fr*)*

[3]*Payame Noor University of Zanjan and Institute of Geophysics, University of Tehran, 14155-6466, Iran*
*(e-mail:* etavabi@yahoo.com*)*

**Abstract.** We analyze multi-wavelength and multi-viewpoint observations of a large-scale event viewed on 7 April 2011 originating from an active region complex. The activity leads to a white-light jet being formed in the outer corona. The topology and evolution of the coronal structures were imaged in high resolution using the *Atmospheric Imaging Assembly* (AIA) onboard the *Solar Dynamics Observatory* (SDO). In addition, large field-of-view images of the corona were obtained using the *Sun Watcher using Active Pixel System detector and Image Processing* (SWAP) telescope onboard the *PRoject for Onboard Autonomy* (PROBA2) microsatellite, providing evidence for the connectivity of the coronal structures with outer coronal features that were imaged with the *Large Angle Spectrometric Coronagraph* (LASCO) C2 on S*olar and Heliospheric Observatory* (SOHO). The data-sets reveal an Eiffel-tower type jet configuration extending into a narrow jet in the outer corona. The event starts from the growth of a dark area in the central part of the structure. The darkening was also observed in projection on the disk by the *Solar TErrestrial RElations Observatory-Ahead* (STEREO-A) spacecraft from a different point of view. We assume that the dark volume in the corona descends from a coronal cavity of a flux rope that moved up higher in the corona but still failed to erupt. The quadrupolar magnetic configuration corresponds to a saddle-like shape of the dark volume and provides a possibility for the plasma to escape along the open field lines into the outer corona, forming the white-light jet**.**

## 1. Introduction

Among the different forms of solar activity, there are apparently linear collimated plasma flows, which are presumably guided by magnetic fields, usually referred to as jets. These are found in the literature concerning soft X-ray and extreme ultraviolet (EUV) observations, and as ray-like features in eclipse observations (Saito and Tandberg-Hanssen, 1973). A wide variety of jet-like structures are observed in the solar atmosphere. They can be formed both from relatively cool plasma as spicules, spikes, macro-spicules and surges (Koutchmy and Stellmacher, 1976; Rompolt and Svestka, 1996; Sterling, 2000; Yamauchi *et al.*, 2005) as well as from hot plasma and seen in X-rays, white-light and EUV (Brueckner, 1981; Shibata *et al.*, 1992; Shimojo *et al.*, 1996; Koutchmy *et al.*, 1997; Koutchmy *et al.*, 1998, 2010; Pike and Mason, 1998; Wang *et al.*, 1998; Chae *et al.*, 1999; Harrison *et al*., 2001). Wang *et al.* (1998) discovered that numerous white-light LASCO-C2 jet-like ejections above the polar regions were the outward extensions of EUV jets seen with the SOHO/*Extreme ultraviolet Imaging Telescope* (EIT), which in turn originated from flaring bright points inside the polar coronal holes, during the 1996 – 1997 activity minimum. Coronal-hole regions still produce white-light jets at solar maximum. However, many ejections are recurrent in nature and originate from active regions located inside or near the boundaries of non-polar coronal holes (Wang and Sheeley, 2002).

In general the topology of a jet resembles the geometry of field lines in the vicinity of a null point. Such magnetic configuration which includes onion-shape separatrices was proposed by Koutchmy et al. 1994 to explain the form of large jets and streamers with a helmet. Observations often show that a smaller size jet may start from the brightening and expansion of a loop within a dome-like magnetic configuration. Shibata *et al*. (1994) described a very strong X-ray jet formation in such a configuration using the *Yohkoh* observations, calling such a structure an anemone active region. The presence of the neutral point naturally suggests that field line reconnection and magnetic field annihilation could occur in it. Shibata *et al*. (1994) came to the conclusion that a magnetic reconnection mechanism is responsible for the large X-ray jet production based on 2D numerical simulations. Later, Shibata (1998, 1999) developed the model in 3D geometry that he called a plasmoid-induced reconnection model. Helically twisted flux ropes play a significant role in this scenario. However the source of energy for the jet formation is magnetic energy release in the reconnection site.

Filippov *et al*. (2009) proposed a qualitative model of jet formation within the "Eiffel tower" magnetic configuration in which the energy source of the process was not energy release in a reconnection site but free magnetic energy of a small twisted flux tube. One of the reasons to look for an alternative mechanism of jet formation was the observed direction of the displacement of coronal loops during the event. Usually, coronal loops are associated with thin magnetic flux tubes filled with denser or hotter plasma than the surroundings. The dynamics of the coronal loops reflects changes in the magnetic field structure. Loops in the events studied by Filippov *et al*. (2009) did not come close to the assumed null point position from opposite directions of two diametrically opposite sides of the saddle structure, as expected in the classical scenario of reconnection; instead they moved in the same direction (one approached to the center of the structure, while the other moved away from it). Filippov *et al*. (2009) suggested that plasma was heated in a small twisted flux tube and confined in a magnetic trap. Coming to the reconnection site near a null point, where the magnetic field is getting weak, the trap is opened and the hot plasma is readily able to move upward along open field lines. The gas pressure gradient works like a piston that pushes plasma through a nozzle.

Shimojo *et al*. (2007) found that the loop expansions occurred on the footpoint of the jets and the jets started just after breaking the expanding loops. The small filament eruption at the footpoint of the EUV-jet was reported by Chae *et al.* (1999). Helically twisted flux ropes play a significant role in the scenario of Shibata (1998, 1999) describing a plasmoid-induced reconnection. Evidence of an helical twist in a EUV polar jet was found by Patsourakos *et al.* (2008) on the basis of the twin STEREO spacecraft observations. They found that the jet's body appeared to untwist while rising and proposed the magnetic untwisting as a driving mechanism for the initiation of polar jets.

In this article, we analyze new multi-wavelength and multi-viewpoint observations of a large-scale event within a complex of active regions that results in a white-light jet formation. Data from several space-born instruments are used, namely the *Large Angle Spectrometric Coronagraph* (LASCO) on the *Solar and Heliospheric Observatory* (SOHO) (Brueckner *et al*., 1995), the Atmospheric Imaging Assembly (AIA) on the Solar Dynamic Observatory (SDO) (Lemen *et al*., 2011), the *Extreme UltraViolet Imager* (EUVI) from the *Sun Earth Connection Coronal and Heliospheric Investigation* (SECCHI) instrument package on the *Solar TErrestrial*

*RElations Observatory* (STEREO) (Wülser *et al*., 2004), the *Sun Watcher using Active Pixel System detector and Image Processing* (SWAP) on PROBA2 (*PRoject for Onboard Autonomy*) micro-satellite (Berghmans *et al*., 2006; Halain *et al*., 2010), as well as groundbased observations from the Big Bear Solar Observatory. These multi-wavelength and multi-spacecraft observations allow us to see now some more details about the 3D geometry and even about the temperature structure in the jet source region.

We then conclude from this analysis that a confined eruption of a twisted rope is an important ingredient in jet formation.

**2. Observations**

NOAA active region 1183 was close to the west limb on 7 April 2011 when some activity occurred in the surrounding corona, leading to the sudden appearance of a narrow bright white-light coronal ray or jet above in the field of view of the LASCO-C2 coronagraph (Figure 1). The whole event lasted for about six hours and showed an outwardly directed proper motion (as manifested by the apparent outward propagation of the intensity enhancement) of about 200 km s$^{-1}$). It is noticeable that the new jet is not a CME and was not even catalogued as an event with the CACTUS rather sensitive data base system recording dynamical coronal phenomena observed with the SOHO/LASCO coronagraphs (*e.g.*, Robbrecht *et al.*, 2009). Accordingly, such long ray or jet coronal structures are similar to what has been described in the past by eclipse observers using white-light pictures (*e.g.* Koutchmy, 2004) and identified as the result of dynamical phenomena occurring above active regions (Slemzin *et al.*, 2008, Koutchmy *et al.*, 2010). The multi-wavelength and multi-viewpoint observations allow us to obtain a more detailed information about the 3D geometry and about the temperature structure in the jet source region.

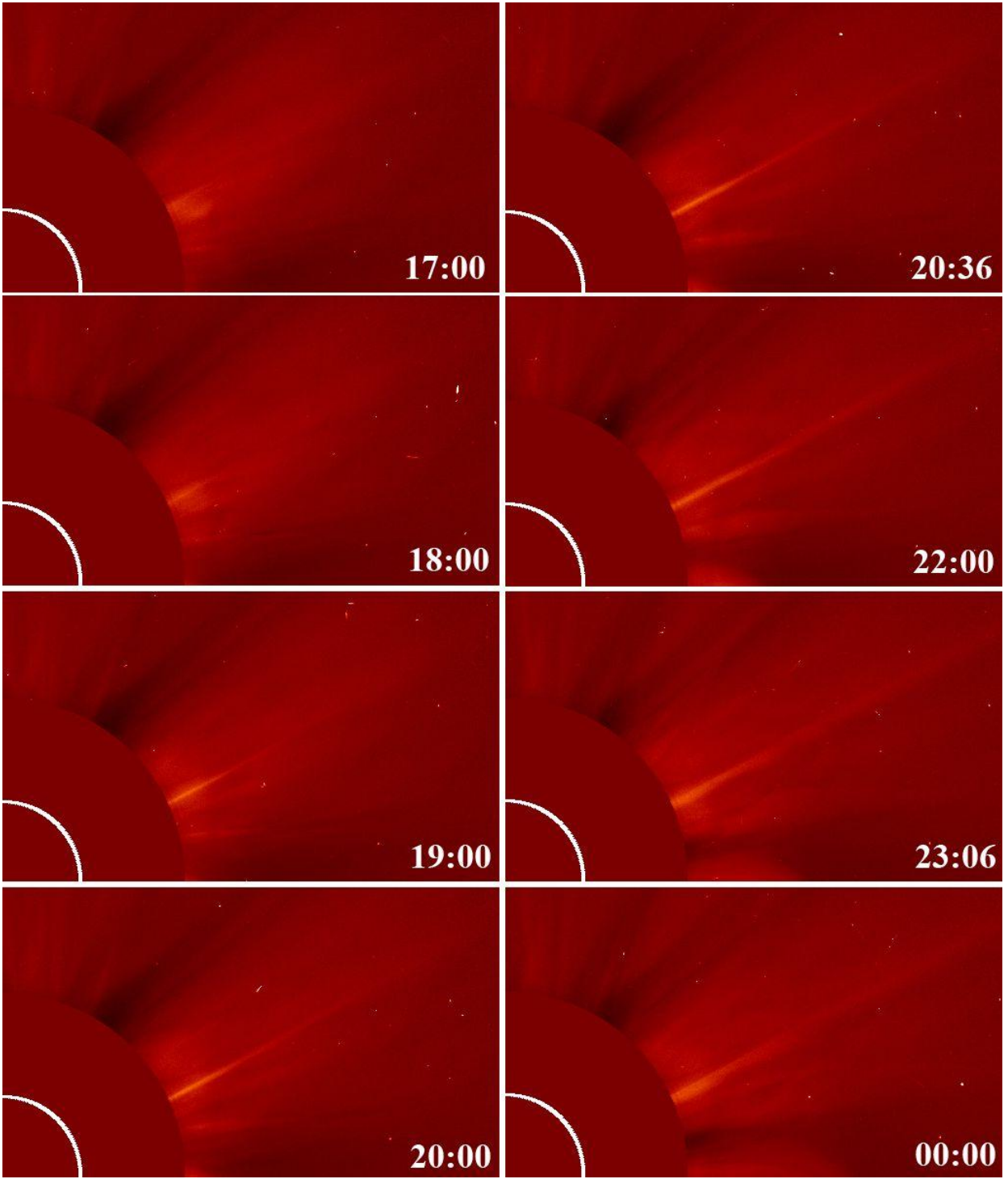

**Figure 1.** Growth of the white-light linear ray like jet in the field of view of the SOHO/LASCO-C2 coronagraph on 7 April 2011. (Courtesy of the SOHO/LASCO Consortium, ESA and NASA).

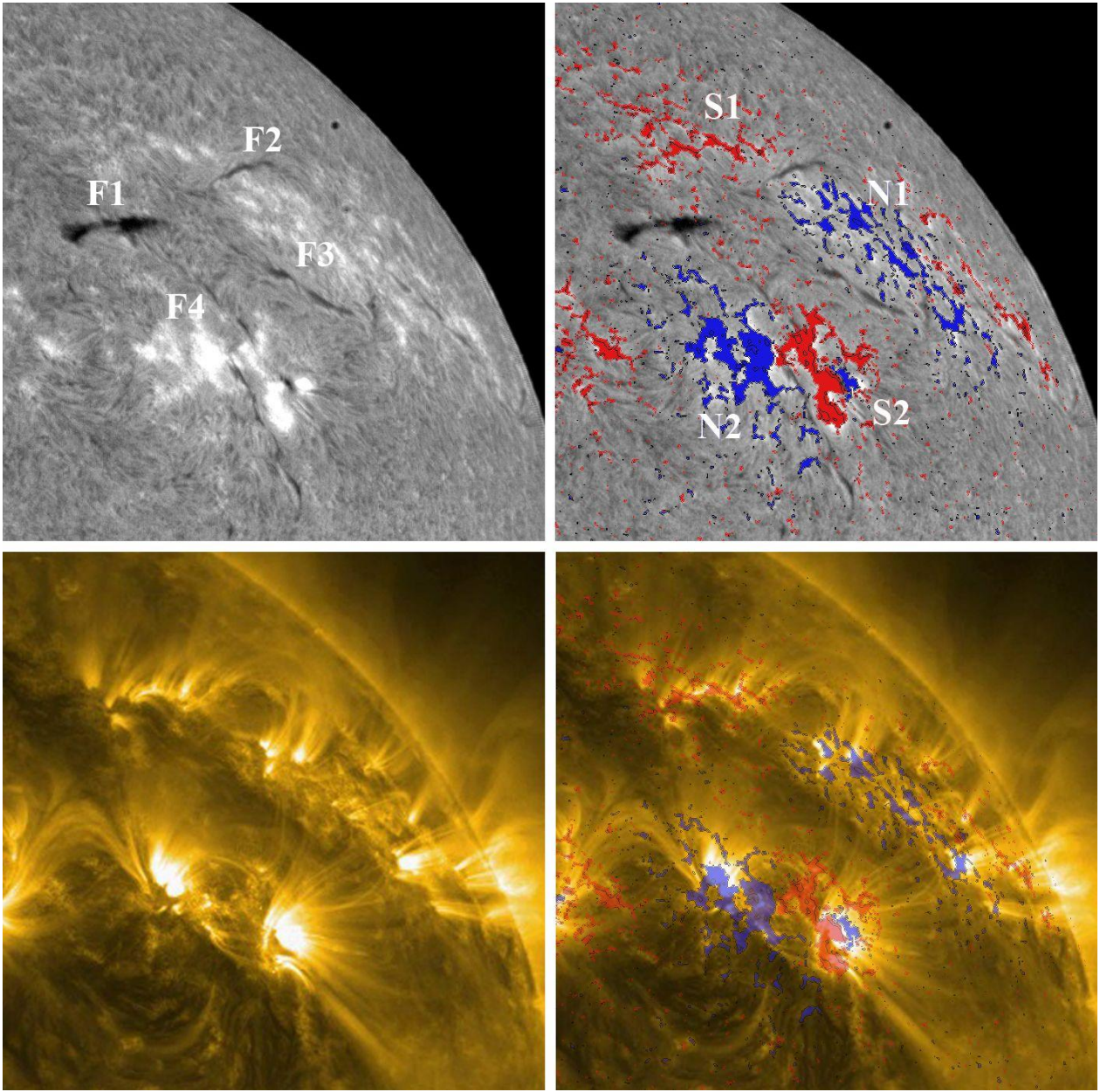

**Figure 2.** Hα (top left) and AIA 171 Å (bottom left) images in the vicinity of AR 1183 on 4 April 2011 at 18 UT. In the right column, the photospheric magnetic-field concentrations stronger than ±50 gauss (G) from the HMI magnetogram are overlapped on the images from the left column. Red areas represent negative polarity, while blue areas represent positive polarity (Courtesy of the Big Bear Solar Observatory, NASA/SDO, and the AIA and HMI science teams).

The active region NOAA 1183, the "dispersed" bipole to the North of it, NOAA 1180, and NOAA 1184 form an intricate active region complex. The whole magnetic structure is very complicated but we believe that only some part of it is related to the jet formation. Figure 2 shows the Hα and EUV images of the central part of the complex region on 4 April 2011 when it was at a more favorable position to look at the surface of the photosphere and chromosphere. In the right column, the photospheric magnetic field concentrations are overlapped on the images from the left column. Four short filaments form a cruciform structure (Filippov, 2011; Filippov and Srivastava, 2011). They are located above the polarity inversion lines (PILs) that separate four magnetic concentrations which jointly form a quadrupolar magnetic configuration of the photospheric magnetic fields. Polarities form two pairs, N1-S1 and N2-S2, which are closely connected by a system of bright coronal loops. However there are also loops connecting the sources N1 to S2, as well as connecting N2 to S1. The bipole N2-S2 belongs to the active region NOAA 1183, while the bipole N1-S1 is formed by a large area of dispersed negative polarity and the following polarity of the decaying active region NOAA 1180. From the flux conservation condition and the continuity of the magnetic field lines, a null point should exist at some height in the center of the whole structure.

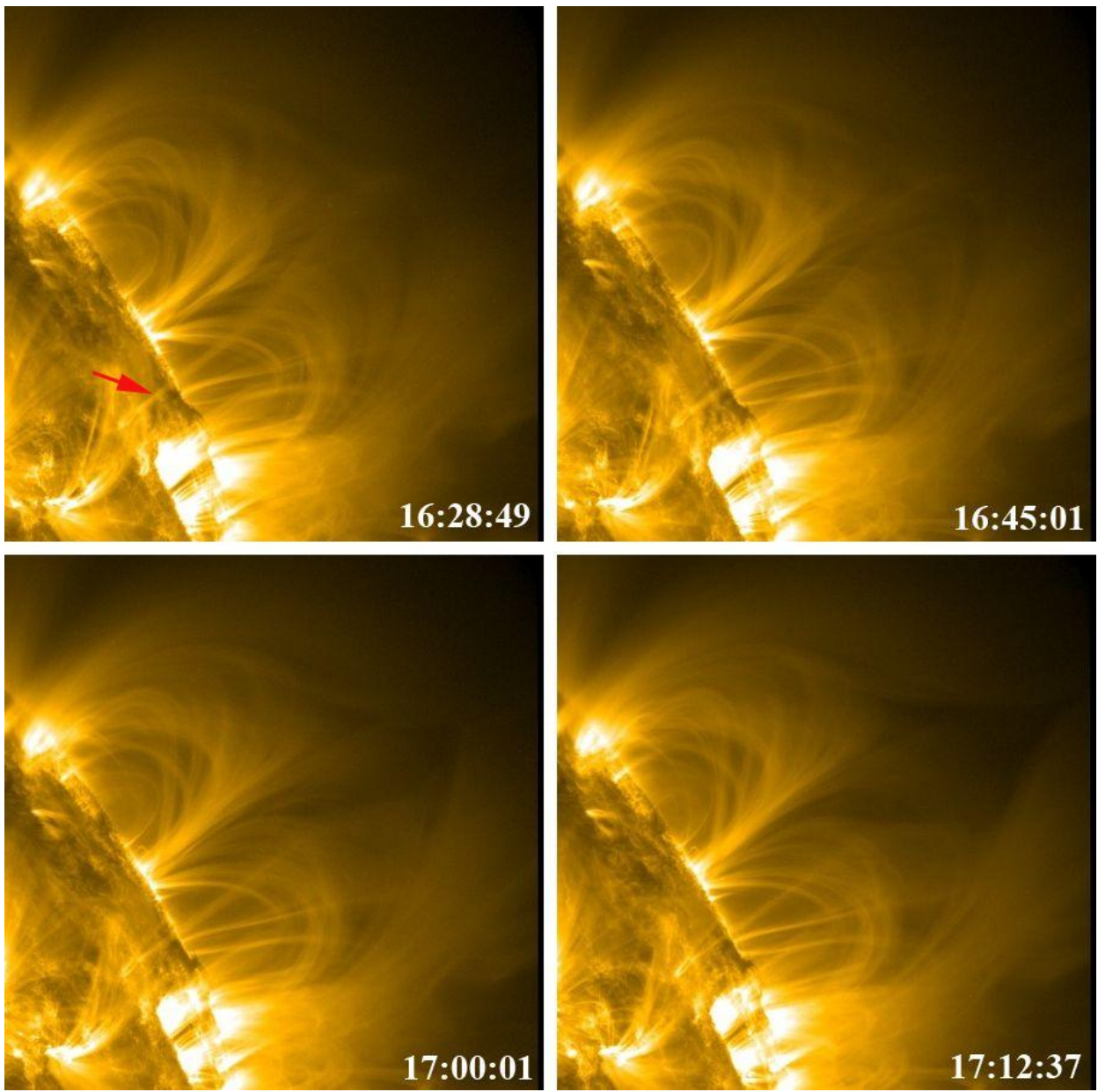

**Figure 3.** AIA 171 Å images showing the development an hyperbolic cavity on 7 April 2011. (Courtesy of NASA/SDO and the AIA science team).

In SDO AIA 171 Å and especially in the larger field of view SWAP images obtained after 16 UT on 7 April 2011 (Figure 3 and Figure 4), the growth of a dark area is seen above the two loop systems. The AIA 171 Å channel shows coronal plasma with rather moderate temperature of 6.3 $\times 10^5$ K. In projection on the sky plane, the dark structure has a saddle-like shape and at larger scale (Figure 4), it is a part of the so-called Eiffel-tower configuration. In other words, the linear white-light jet is the long outward extension of the Eiffel-tower structure (Figure 5). In addition the sequence of high-resolution 171 Å images (see movie 1) shows the displacement of several bright coronal loops during the event. High loops belonging to the bottom loop system S2-N2 first increase in size and then move to the right and down like if they were pushed away from the center of the dark structure. More clearly the evolution of an individual loop can be evaluated like it is for the thin separate loop indicated in Figure 3 by the red arrow. Loops in the top system S1-N1 move towards the solar surface and away from the center of the structure at the beginning of the event. No additional brightening was seen in the 171 Å channels except a rather low loop arcade, which appeared near 18:30 UT, after the formation of the dark saddle structure. Loops connecting S2 and N1 seem to be undisturbed by the event, while loops connecting S1 and N2 can be hardly recognized in the images. Diffuse loops that constitute the upper border of the dark area form a cusp of an Eiffel-tower structure moving up and aside from each other. They apparently are rooted at the negative polarity concentrations S1 and S2.

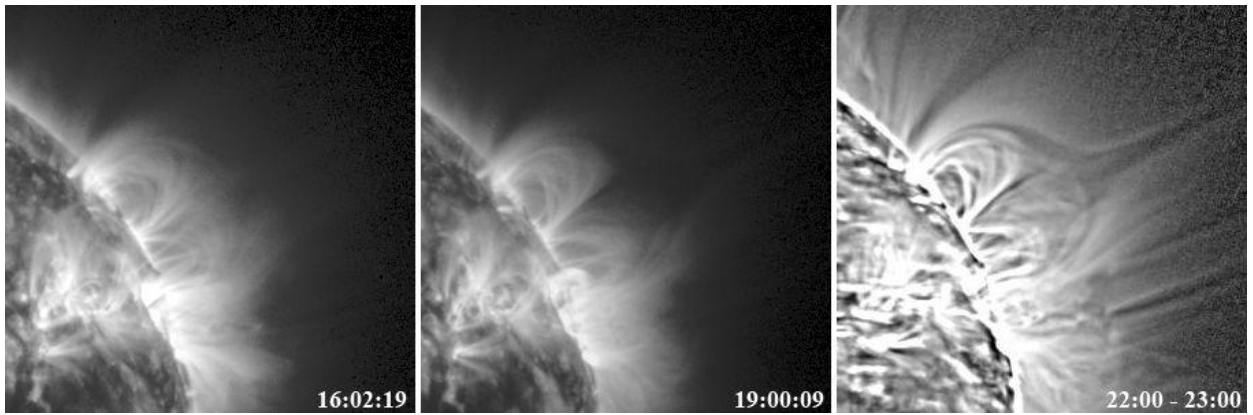

**Figure 4.** SWAP 174 Å images showing the development of the hyperbolic cavity to large distance from the limb. The right-hand image is obtained by summing 200 images taken from 22 UT to 23 UT and applying the unsharp mask filter after. (Courtesy of the PROBA2 team. SWAP is a project of the Centre Spatial de Liège and the Royal Observatory of Belgium funded by the Belgian Federal Science Policy Office (BELSPO)).

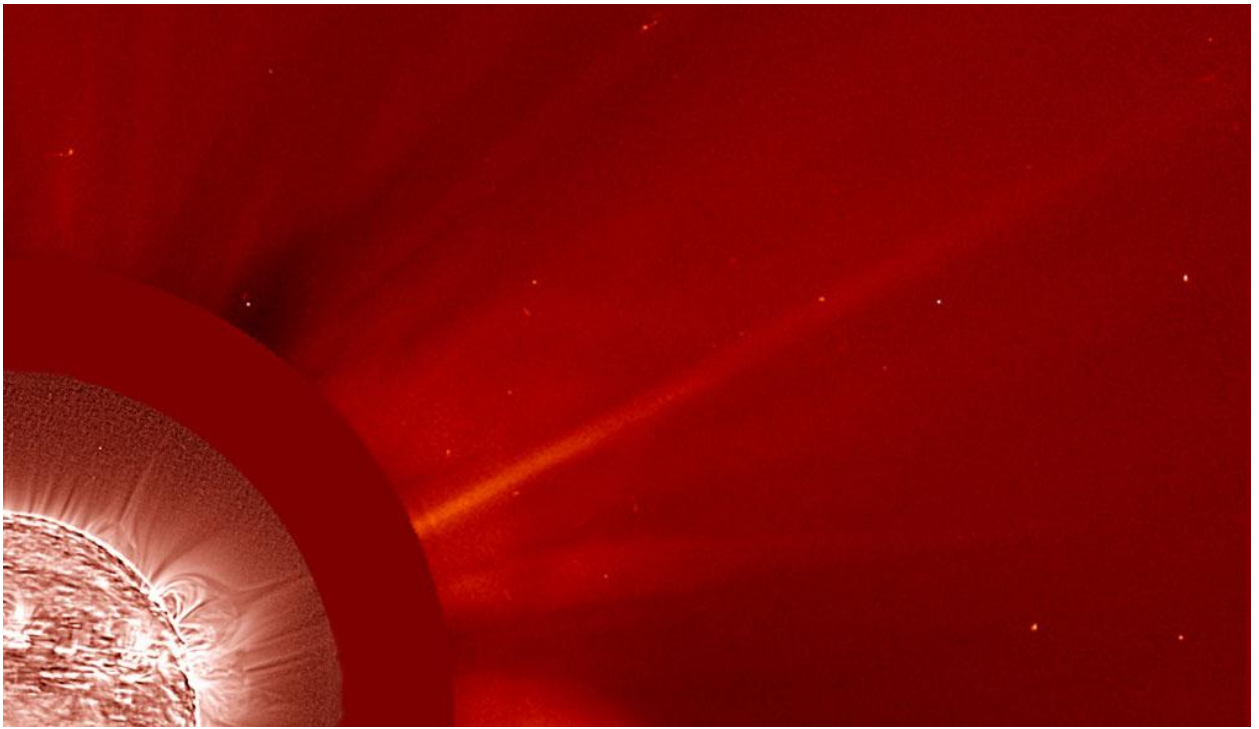

**Figure 5.** Composite image of the white-light SOHO/LASCO-C2 image on 7 April 2011 at 22:00 UT and SWAP 174 Å image (inside) obtained by summing 200 images taken from 22 UT to 23 UT and applying the unsharp-mask filter after.

Meanwhile the AIA 94 Å channel which is showing plasma ten times hotter ($6.3 \times 10^6$ K) than the 171 Å channel ($6.3 \times 10^5$ K), reveals quite different brightness changes (Figure 6). The event starts at 16 UT with the appearance of bright low loops nearly at the same place where the bright loops were observed in the 171 Å channel later at 18:30 UT (see movie 2). Simultaneously, rising bright loops appear closer to the central part of the quadrupolar loop system. These loops could belong to the loop system S2-N1 or S2-N2. The rising motion stopped near 17 UT at a rather limited height of order of 70 Mm. The brightening at the top of the loops spreads out after that and fades away. Synchronously with the rising of the bright loops, a dark area spreads above them and shows higher loops. In the 94 Å channel the event looks like a failed eruption or a confined flare.

The STEREO-B spacecraft observed the active complex region of interest on the disk close to the central meridian. Figure 7 presents a series of STEREO-B/SECCHI/EUVI 195 Å images of the jet source region on 7 April 2011 (see also movie 3). An unsharp mask filter was applied to make individual coronal loops more visible. Three loop systems S1-N1, S2-N1, and S2-N2 are seen very clearly, while the system S2-N1 is fainter but also recognizable in the images. A dark structure in form of a cross is located in the center of the active region complex. We assume it is a characteristic quadrupolar filament channel corresponding to the four filaments shown in Figure 2. Prominent changes appeared at 16:45 UT when a thin bright loop arises between S2 and N2 polarities. At the same time tiny brightenings appear on both sides of the polarity inversion lines (PILs) separating S2 and N2, as well as for S1 and N2 above which the filaments F1 and F4 are respectively located. They delineate the dark cross. Later on, a small (low) arcade of bright loops arises connecting the brightening on both sides of the southern branch of the PIL. No such arcade appears above the eastern part of the PIL, although some thin features can be vaguely recognized along the PIL. The thin bright loop to the south of the cross slightly increases in size later and fades out near 17:30 UT. After 18:30 UT a new very bright growing loop system arises at the same place.

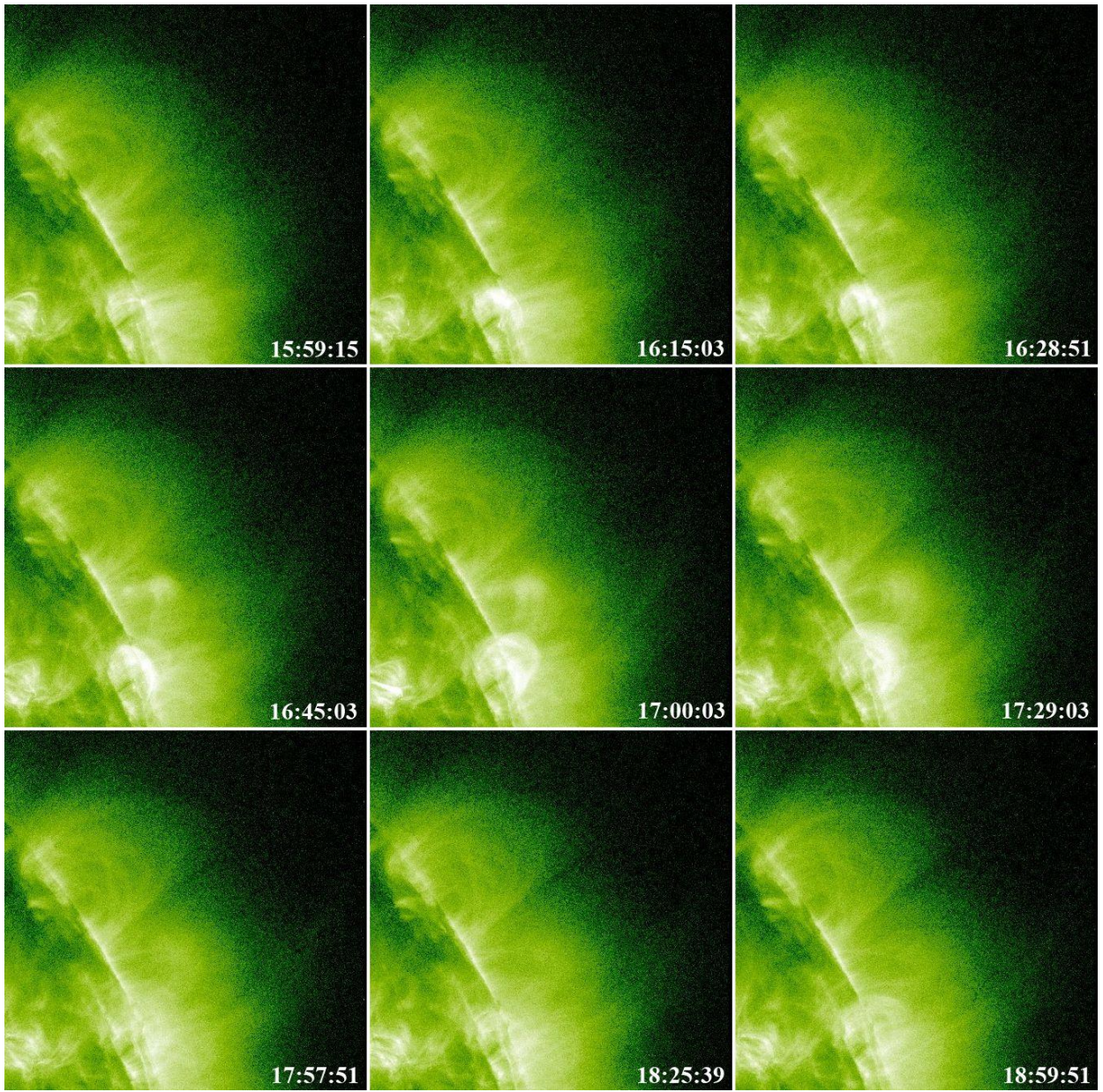

**Figure 6.** April 7, 2011 AIA 94 Å images showing the raising of a bright loop system below the hyperbolic cavity. (Courtesy of NASA/SDO and the AIA science team).

There are also different changes in the loop structure of the active complex which are more directly related to the formation of the saddle-like dark structure that was observed on the limb by the Earth-orbiting spacecraft SWAP and SDO. The first one is the disappearance of a bright feature at the southern end of the cross between 16:30 UT and 16:45 UT. This feature is indicated by the green arrow in Figure 8. This is possibly the brightest legs of arches above the southern part of the PIL. The second one is the displacement of the loops connecting S1 and N1 and passing through the central area of the active complex away from the center in the direction shown by the red arrow in Figure 9. The faint loops connecting S1 and N2 seem also to move away from the center of the complex. Thus, soon, by 18:00 UT, the central area of the complex becomes free of bright loops and therefore dark.

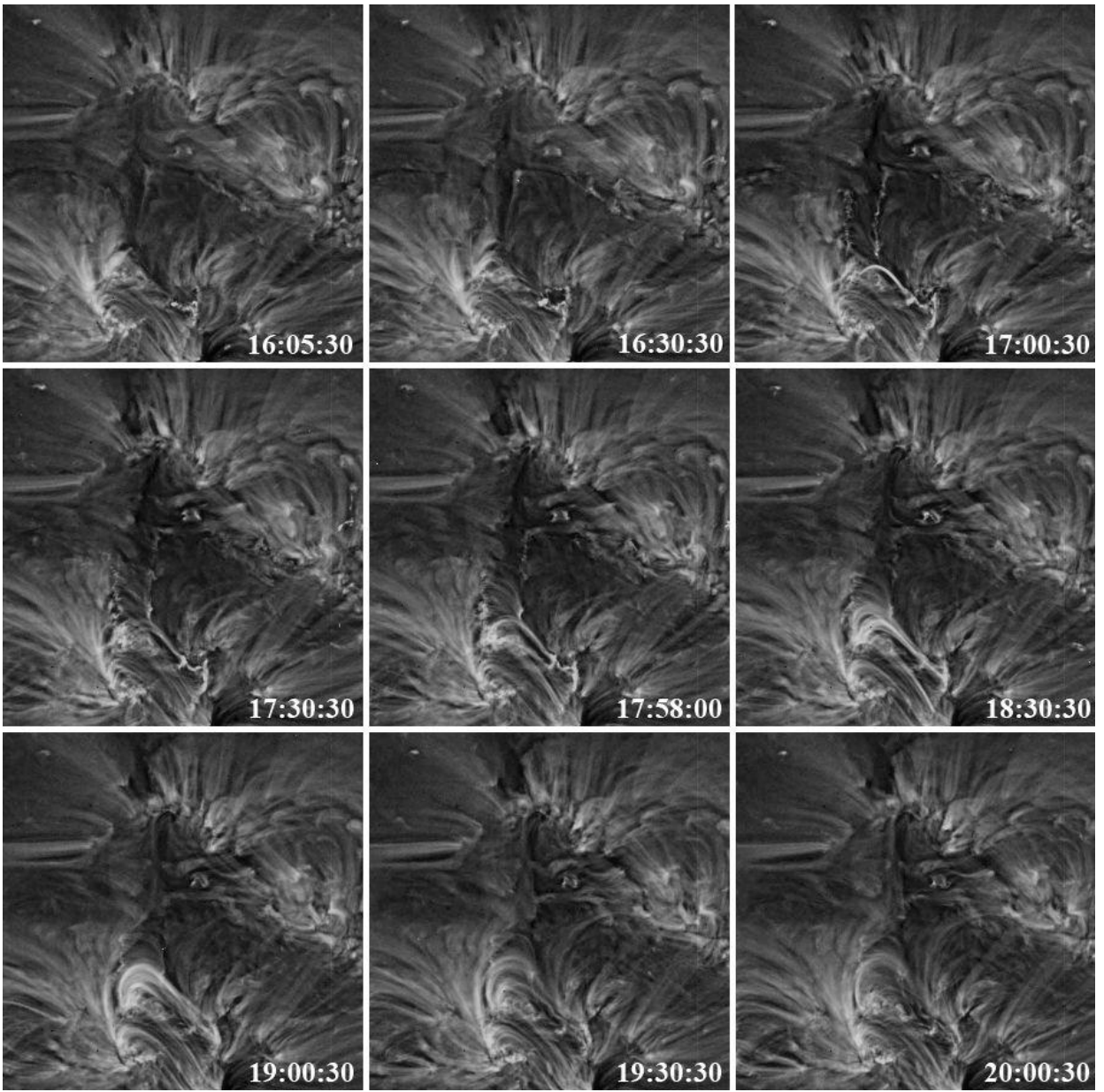

**Figure 7.** STEREO-B/SECCHI/EUVI 195 Å images of the jet source region on 7 April 2011. An unsharp mask filter was applied to make individual coronal loops more visible. (Courtesy of STEREO/SECCHI Consortium).

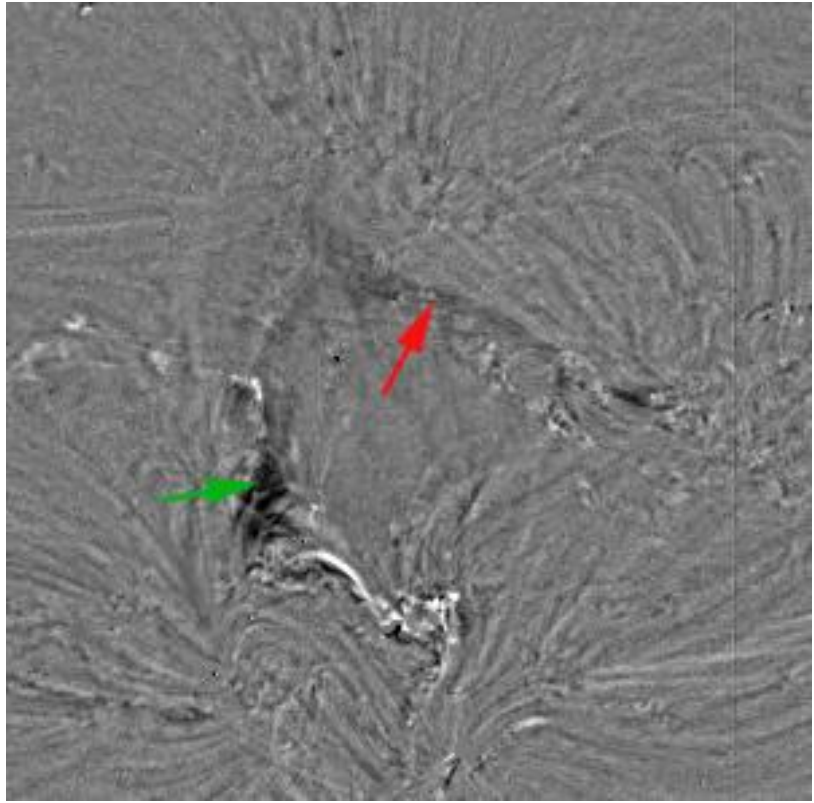

**Figure 8.** STEREO-B/SECCHI/EUVI 195 Å difference image from the frames taken at 16:45 UT and the frame taken at 16:30 UT.

## 3. Discussion and Scenario of the Event

Comparing the time of appearance of the jet in the field of view of LASCO-C2 and the time of formation of the dark saddle structure in the EUV images, as well as comparing their position angles, convincingly show that the former results from effect of the latter. The geometry of the jet source region, as it is concluded from the limb observations, looks very typical for many jet models. It is the so-called Eiffel tower or sea-anemone configuration (Shibata Nozawa, and Matsumoto, 1992; Shibata *et al.*, 1994; Zirin and Cameron, 1998, Filippov *et al.*, 2009). It can be created by a patch of parasitic polarity inside a large unipolar cell. In planar 2D geometry, at least three magnetic poles should exist for the creation of a null point within the configuration.

Figure 9 shows a schema of the magnetic configuration of the active region complex, that we deduce from the set of observational data presented in the previous section. Two pairs of large scale bipoles located close to each other form a quadrupole. Negative polarity (red in Figure2) seems to be prevailing because the S1 and S2 concentrations are connected by dispersed negative elements. That is why some field lines from the peripheries of the negative polarities can be open. The presence of the filaments indicates the existence of flux ropes within the configuration. According to the shape of the photospheric PILs separating the opposite polarities, there could be two flux ropes, FR1 and FR2, curved nearly at a right angle. The fine structure of the filaments allows us to conclude that the chirality of all filaments is dextral, typical for filaments in the northern hemisphere. Therefore the flux ropes should have a negative helicity and their field lines have the shape of left helices.

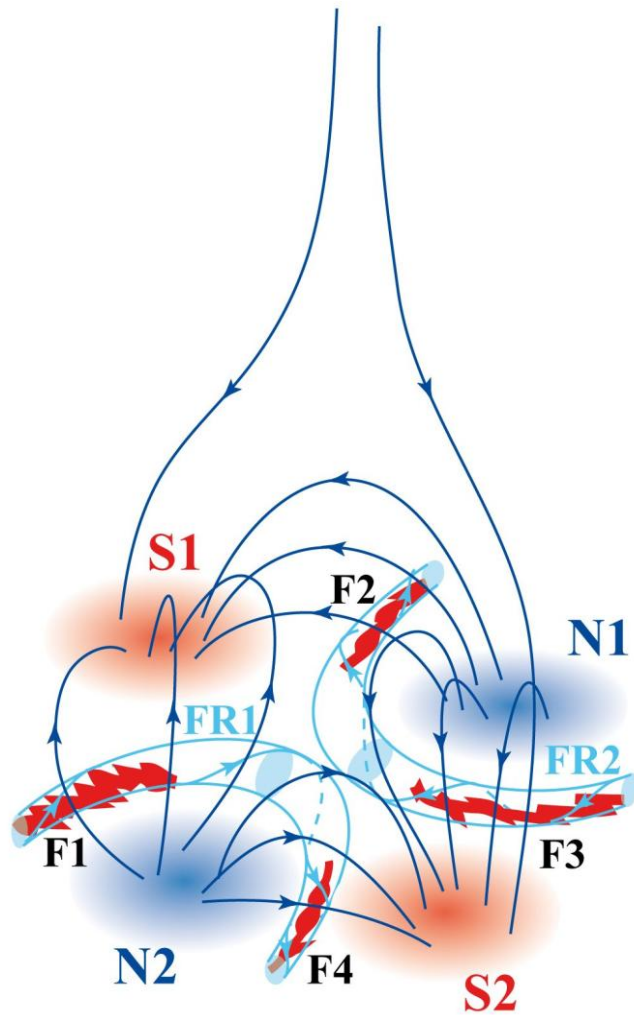

**Figure 9.** Schema of the Magnetic Configuration of the Active Complex before the Event.

The beginning of the event resembles a failed or confined eruption very much (Filippov and Koutchmy, 2002). The first manifestation is seen in the Fe XVIII 94 Å line as the rising of a bright arch, which stops at the height of 70 Mm. Since two flux ropes are present in the complex, it is natural to associate the ascending of the bright structure with the instability of one of them. STEREO observations show that it should be FR1, because the elements of the flare ribbons and the arcade appear along the PIL after. The upper parts of a flux rope are usually made of more rarefied plasma and form a coronal cavity (Harvey 2001; Gibson *et al.,* 2006). During an eruption, the dark cavity increases in size and moves faster than the central part of the flux rope (Filippov, 1996). As a rule, a failed eruption is not followed in its upper part by a coronal manifestation such as a CME. If the eruption happened in a simple bipolar arcade, we would possibly observe nothing above in the C2 coronagraph field of view. But the event took place inside an active complex with a complicated quadrupolar structure. The feet of the flux rope FR1 are rooted in stronger fields, especially the southern end located within the S2-N2 bipolar region. The middle part of the FR1 is situated in a weaker field region in the vicinity of the null point. Tying action of the ambient field on the flux rope is less here than at the ends, therefore the middle is able to rise to a higher altitude. Comparison of Figure 9 with Figure 10 shows that our schema is in a good agreement with the calculated potential magnetic field structure. Most of the open field lines emanate from a negative polarity. The whole configuration has a significantly 3D shape without any preferable dimension. It is unlikely that such configuration contains an elongated current sheet like in a popular streamer model.

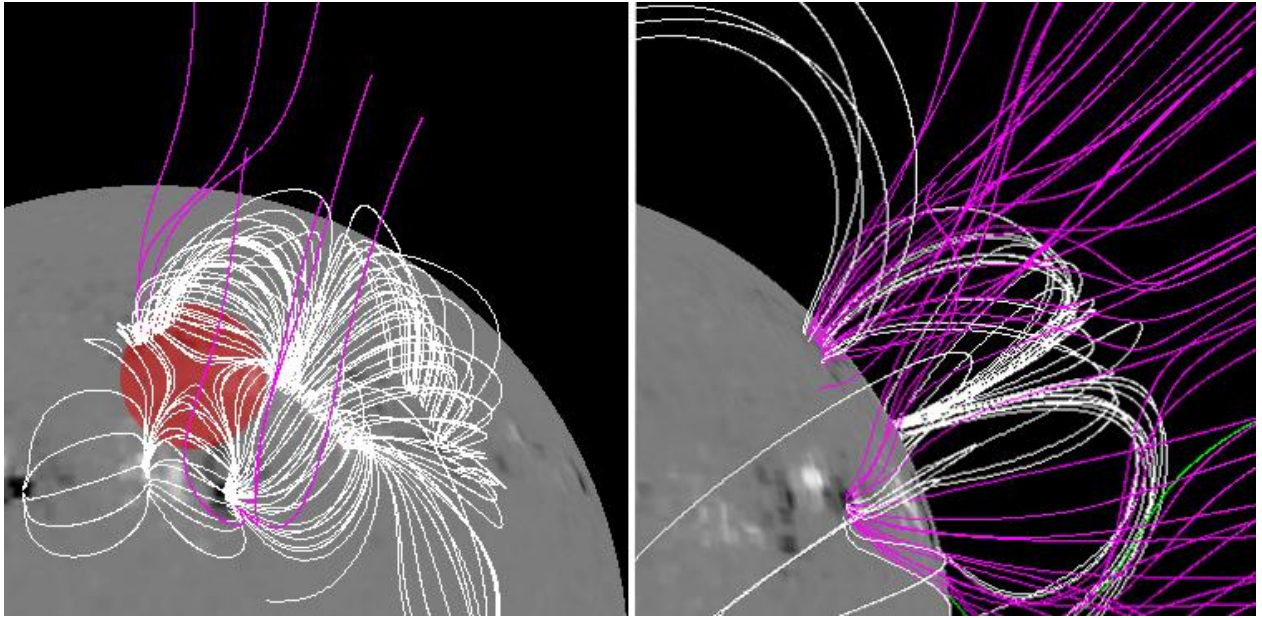

**Figure 10.** Potential magnetic field lines of the region of interest on 4 April 2011 (left) and 7 April 2011 (right) calculated using a PFSS model. Pink lines show open field lines emanating from a negative polarity. Red circle in the left panel shows the central part of the photospheric quadrupole. (Courtesy of Lockheed Martin Solar and Astrophysics Lab)

We believe that the dark cavity of FR1 rises to a higher altitude during the failed (confined) eruption of FR1 and occupies the vicinity of the coronal null point. The magnetic field and the plasma pressure of FR1 push out the field flux tubes belonging to the S1-N1 and to the S2-N2 systems, which leads to the creation of the large dark saddle-shaped volume. Reconnections could take place in this configuration. However, we do not see obvious signs of the reconnection processes. After the first rise of the bright arch in the Fe XVIII 94 Å line to the height of 70 Mm, all coronal loops look like pushing out the center of the dark structure, at the height of about 190 Mm, both from the limb and from the disk images. It is especially evident in the AIA Fe IX 171 Å images. "Open" flux tubes, which bound the dark structure from above, move up and away from each other instead of approaching to one of them toward the reconnection site. We may guess that the plasma pressure plays a significant role in the event. Although the density is reduced in the flux-rope cavity but the temperature is not, after rising to a height several times greater than the initial value, the pressure within the cavity can exceed the pressure of the surrounding plasma. Somehow this plasma fills the volume surrounding the null point and pushes out the neighboring field lines. Asymptotes of the saddle are the easiest ways for the plasma to spread out along them, and one of the asymptotes turns out to be an open field line. We propose that the coronal plasma flows from the dark saddle-like volume outward along the open asymptote of the saddle and forms the white-light jet visible after in the LASCO-C2 images.

Thus, the energy source responsible for the formation of the white-light jet is the free magnetic energy of the flux rope. Its failed eruption results in a plasma pressure enhancement within the saddle-like magnetic configuration around the null point. The plasma pressure accelerates the plasma flow along the open asymptote of the saddle, forming the rather slow white-light jet becoming a long linear ray.

The proposed scenario is similar to the model suggested by Filippov *et al*. (2009) for small X-ray jets. In this greatly larger-scale event, we now see the details of the development of the complex phenomena in a much more easy way.

**4. Conclusions**

Multi-wavelength and multi-viewpoint observations of the white-light jet formation seen on 7 April 2011 allow us to see the details of the jet's source region. It is the complex set of active regions that form a large-scale quadrupolar magnetic configuration. At least two coronal flux ropes are present within the region before the event. They are revealed by the set of four filaments. Observations in the 171 Å line show the event starting by the appearance of the dark saddle-like cavity above this complex active region. However, in the 94 Å line we see the bright feature rising from a lower to higher height just before the cavity appearance. It resembles very much to a failed or confined eruption. We suppose that the free magnetic energy stored within the flux rope was not great enough to produce a plasma ejection into the interplanetary space but it was sufficient to make a significant rearrangement of the magnetic field and of the plasma coronal structure above the complex active region. Open field lines at the top of the Eiffel tower configuration produce a collimated beam of plasma above the flux-rope cavity and guide it to form the jet and further out, a linear long ray.

**Acknowledgements**  The authors would like to thank the anonymous referee for helpful comments that have improved the manuscript. This work was supported in part by the GI Program of the PROBA2 mission of Royal Observatory of Belgium (ROB) and in part by the Russian Foundation for Basic Research (grant 09-02-00080). The authors are grateful to the Director of ROB R. Van der Linden for his hospitality, to Anne Vandersyppe who made their stay enjoyable, and to David Berghmans and Anik De Groof for their efficient leadership of the GI program.  We also greatly appreciated the availability of the ESA and NASA solar data from the highly successful SOHO, STEREO, and SDO missions. SWAP is a project of the Centre Spatial de Liège and the Royal Observatory of Belgium funded by the Belgian Federal Science Policy Office (BELSPO).
.